\documentclass[11pt]{article}
\usepackage{graphicx}
\usepackage{amssymb}
\usepackage{amscd}
\usepackage{mathrsfs}
\usepackage{longtable,lscape}
\usepackage{amsthm}
\usepackage{amsfonts}
\usepackage{amsmath}
\usepackage{bbm}
\usepackage{float}
\usepackage{url}
\usepackage{hyperref}
\usepackage[margin=0.5cm,font=footnotesize]{caption}
\usepackage{lineno}
\usepackage[round]{natbib}
\usepackage{appendix}
\usepackage{subfig}

\begin{document}
\title{\bf Identifying Quantum Mechanical Statistics \\ in Italian Corpora}
\author{Diederik Aerts$^*$, Jonito Aerts Arg\"uelles$^*$, Lester Beltran$^*$, \\ Massimiliano Sassoli de Bianchi\footnote{Center Leo Apostel for Interdisciplinary Studies, 
        Vrije Universiteit Brussel (VUB), Pleinlaan 2,
         1050 Brussels, Belgium; email addresses: diraerts@vub.be,jonitoarguelles@gmail.com,lestercc21@yahoo.com,autoricerca@gmail.com} 
        $\,$ and $\,$  Sandro Sozzo\footnote{Department of Humanities and Cultural Heritage (DIUM) and Centre CQSCS, University of Udine, Vicolo Florio 2/b, 33100 Udine, Italy; email address: sandro.sozzo@uniud.it}              }

\date{}

\maketitle
\begin{abstract}
\noindent 
We present a theoretical and empirical investigation of the statistical behaviour of the words in a text produced by human language. To this aim, we analyse the word distribution of various texts of Italian language selected from a specific literary corpus. We firstly generalise a theoretical framework elaborated by ourselves to identify `quantum mechanical statistics' in large-size texts. Then, we show that, in all analysed texts, words distribute according to `Bose--Einstein statistics' and show significant deviations from `Maxwell--Boltzmann statistics'. Next, we introduce an effect of `word randomization' which instead indicates that the difference between the two statistical models is not as pronounced as in the original cases. These results confirm the empirical patterns obtained in texts of English language and strongly indicate that identical words tend to `clump together' as a consequence of their meaning, which can be explained as an effect of `quantum entanglement' produced through a phenomenon of `contextual updating'. More, word randomization can be seen as the linguistic-conceptual equivalent of an increase of temperature which destroys `coherence' and makes classical statistics prevail over quantum statistics. Some insights into the origin of quantum statistics in physics are finally provided.
\end{abstract}

\section{Introduction\label{intro}}
In classical statistical mechanics, the behaviour of a large number of identical systems, or `entities', at thermodynamic equilibrium is governed by `Maxwell--Boltzmann statistics', which arises as a consequence of the `distinguishability' and `independence' that are assumed for classical physical entities and their states, respectively. On the contrary, the received view of quantum statistical mechanics is that identical entities are `completely indistinguishable' from a physical point of view. This is mathematically incorporated in two postulates of quantum mechanics, namely, the `tensor product postulate', which rules the representation of composite entities, and the `symmetry exchange postulate', according to which the state vector of two identical entities has to be either symmetric or anti-symmetric with respect to the exchange of any two entities \citep{eisbergresnick1985,huang1987}. The spin-statistics theorem of quantum field theory then states that entities with integer spin must be described by symmetric states, while entities with semi-integer spin must be described by anti-symmetric states \citep{pauli1940}. 

The symmetry exchange postulate has substantial consequences on the statistical behaviour of identical quantum entities at thermodynamic equilibrium, as it entails that entities with integer spin, e.g., photons, obey `Bose--Einstein statistics', whence the name `bosons' given to them, while entities with semi-integer spin, e.g., electrons, obey `Fermi--Dirac statistics', whence the name `fermions'. The symmetry exchange postulate is also responsible of the lack of statistical independence that characterizes identical quantum entities, because the symmetry or anti-symmetry requirement mathematically forces the states of these entities to be `entangled states', which would be responsible of the lack of statistical independence of single-entity states, or `micro-states'.\footnote{Statistical independence means that the probability that two entities are in a given micro-state is the product of the probabilities that each entity is in that micro-state. Both bosons and fermions violate statistical independence, because bosons tend to be in the same micro-state more frequently than statistical independence would allow, while fermions cannot be in the same micro-state.\label{statind}} More precisely, bosons tend to clump, or cluster, together in the same micro-states, with the `Bose--Einstein condensate' as a limit at low temperatures where all entities are in the same micro-state \citep{cornellwieman2002,ketterle2002}. On the contrary, fermions cannot be in the same micro-state, as established by the Pauli exclusion principle.

The above view, though empirically satisfying \citep{annett2005}, has been questioned both from an epistemological \citep{frenchredhead1988,saunders2003,mullerseevinck2009,krause2010,diekslubberdink2020} and a physical point of view \citep{hong1987,knill2001,zhao2014}. In particular, quantum experimentalists seem to treat photons of different energy as distinguishable entities, while they would be indistinguishable according to quantum mechanics \citep{zhao2014}.

We have recently investigated the statistical behaviour of the words contained in medium- and large-size texts produced by English language, discovering that they obey Bose--Einstein, rather than the Maxwell--Boltzmann, statistics \citep{aertsbeltran2020,beltran2021,aertsbeltran2022a,aertsbeltran2022b,philtransa2023,independence2024}. We have also put forward the hypothesis that `meaning plays in the linguistic-conceptual domain a role that is very similar to the role played by quantum coherence in the physical domain
\citep{aertsbeltran2022a,aertsbeltran2022b,independence2024}. This research fits a bold and novel research programme, called `quantum cognition', 
which applies the mathematical formalism of quantum mechanics in Hilbert space to the modelling of high-level cognitive phenomena, e.g., categorization, language, perception, judgement, and decision-making (see, e.g., the volumes \citet{vanrijsbergen2004,khrennikov2010,busemeyerbruza2012,havenkhrennikov2013,melucci2015,havenkhrennikovrobinson2017} and references therein). And, we have substantially contributed to the development of the quantum cognition programme (see, e.g., \citet{aertsaerts1995,aerts2009a,aertsbroekaertgaborasozzo2013,aertsgaborasozzo2013,aertssassolisozzo2016,aertssozzo2016,aertsetal2018a,pisanosozzo2020,arguellessozzo2020,aertssassolisozzoveloz2021}).

In this article, we present a theoretical and empirical investigation on the statistical distribution of the words contained in selected large-size texts written by Italian authors of XIX and XX century and collected in a digital library of Italian language. To this end, we review and generalize in Sect.~\ref{theory} a theoretical framework elaborated by ourselves for the identification of quantum statistical structures in the linguistic-conceptual domain. In this framework, defined `energy levels' are attributed to the words of a text according their decreasing ranking, i.e. the most frequent word in the text is attributed the lowest energy level. Then, we analyse in Sect.~\ref{experiment} the empirical data collected on the number of appearance of the words of the above literary texts of Italian language, and demonstrate that, in all texts, their words distribute according to Bose--Einstein statistics, showing at the same time significant deviations from Maxwell--Boltzmann statistics. This leads us to conclude that, as in the case of English language, the words of an Italian text behave as an `ideal gas of cognitons in thermodynamic equilibrium', where the term `cogniton' has been introduced in \citet{aertsbeltran2020} to denote the fundamental quantum of human language, photons being the fundamental quanta of electromagnetic radiation. 

Then, we present in Sect.~\ref{randomization} the results of a variant of the empirical study in Sect.~\ref{experiment}, in which we introduce an effect of `word randomization' on the texts of Italian language studied in that section. More specifically, the words in the text are selected randomly this time, drawing them from the collection of words that make up the original text, thus changing their number of appearance across the different placeholders. We demonstrate that, in all randomized texts, while the Bose--Einstein statistics still provides a good model for the distribution of the randomized text, the Maxwell--Boltzmann statistics starts to be a good model too. This can be explained by hypothesizing that randomization of words produces on the meaning coherence of a text an effect which is the linguistic-conceptual counterpart of the decoherence effect produced by an increase of the temperature in a quantum gas \citep{independence2024}.

Next, we propose in Sect.~\ref{explanation} an explanatory hypothesis on the role that meaning might play in human language. Indeed, the words of a text carry meaning and, each time a word is added to a text, it has to preserve the meaning coherence of the overall text and, at the same time, the word `contextually updates' that overall meaning. Using the terminology that is typical of quantum mechanics, the word, and the concept corresponding to that word, has to entangle with all the other words that are already present in the text. This phenomenon of `entanglement formation through contextual updating' is such that the words that have more affinity with the overall meaning of the text also have a higher probability of occurrence in the new text, which determines the lack of statistical independence that is also typically observed in human language. In particular, identical words, because they have the same meaning, tend to clump, or cluster, together, as expected within Bose--Einstein statistics. More important, the appearance of Bose--Einstein statistics is only partially due to the conceptual indistinguishability of identical words. Indeed, while identical words are conceptually indistinguishable, different words are not conceptually indistinguishable. In general, what is really important for the identification of a quantum statistical behaviour is the existence of an abstract conceptual level, rather than a concrete objectual level, for the entities in question, in our case the words that appear in a text. Indeed, while two identical concepts in a piece of text can be interchanged without changing its overall meaning, this exchange invariance no longer applies to objects.

Finally, we provide in Sect.~\ref{physics} some insights about the appearance of quantum mechanical statistics in physics. More precisely, we suggest that, what we have hypothesized above on the presence of Bose--Einstein statistics in linguistic-conceptual domains, is exactly what might occur in physics. This would align with a position maintained by Einstein on the appearance of the Bose--Einstein statistics in physics before the  advent of quantum mechanics. Moreover, this would provide a physically more founded justification than the symmetry exchange postulate for the lack of statistical independence that is observed in physics. Finally, our hypothesis would dissolve the discrepancy mentioned at the beginning of this section between how theoretical physicists and experimental physicists look at indistinguishability of identical photons.

To conclude, the results obtained in the present article on texts of Italian language confirm and strengthen those obtained on texts of English language \citep{aertsbeltran2020,beltran2021,aertsbeltran2022a,aertsbeltran2022b}, which suggests that there are underlying mechanisms that are meaning-related and are independent of the specific language that is used to express that meaning. 
Furthermore, these results (i) naturally fit a novel interpretation of quantum mechanics, the `conceptuality interpretation', initiated by one of us, which regards physical entities as concepts rather than physical objects, and (ii) suggest a new theoretical direction for the development of a quantum thermodynamic theory, as mentioned in Sect.~\ref{conclusion}.

\section{A theoretical framework for the statistical distribution of the words in a text\label{theory}}
We present in this section a general theoretical framework for the identification of quantum statistical structures in texts produced by human language. The framework, though relying on our recent investigation on these structures, can be considered as a deepening and extension of it \citep{aertsbeltran2020,aertsbeltran2022a,aertsbeltran2022b,independence2024}. Before proceeding, however, we preliminarily need to introduce some basic linguistic notions and terminology, as follows \citep{aerts2009a,aertsgaborasozzo2013,aertssassolisozzo2016,aertssozzo2016,pisanosozzo2020,aertssassolisozzoveloz2021,aertsbeltran2020}.

In the theoretical linguistic-conceptual framework we put forward, `words' are names, or labels, for `concepts'. A concept is an abstract entity that we assume to be always in a defined `state'.\footnote{The introduction of concepts as abstract entities may suggest some analogies with other philosophical theories of concepts (see, e.g., \citet{peacocke1992}). We stress, however, that we do not consider the terms `concept' and `meaning' as synonyms in our framework. We can best note that this is the case in human language by considering the combination of different concepts and noting how the meaning of such a combination usually comes about only when enough concepts of the combination have been reified and often even before they have all been reified. In this sense, `meaning' is a notion that is carried in complex ways and potentially rather than already actually by individual concepts, and it is the overall mechanism that takes place during the combination of concepts, including the entanglement that occurs, that produces meaning. In our comparison of human language with quantum structures, we place meaning in terms of human language where quantum coherence is in terms of interacting quantum entities. Elaborating further on these aspects would distract us too far from the topic of this article and we refer the interested reader to our past publications (see, e.g., \citet{aertssassolisozzo2016,ijtp2023,ass2025a,ass2025b}).} The state of a concept captures the `meaning' carried by the concept and can change under the influence of a `context'. Individual concepts can then be combined, or composed, to form more complex conceptual entities, as `conceptual combinations'. Concepts and their combinations are examples of general `entities of meaning'. Next, a `text' produced by human language, e.g., a story-telling text, is a complex entity of meaning made up of the concepts corresponding to all the words that appear in the text.

Let us now consider a text ${\cal T}$ made up of $N$ words in total, of which $n\leq N$ are different from each other. Let us order these different 
words according to their decreasing `number of appearance' in ${\cal T}$. Let us denote the different words, ordered in this way, by $w_0$, $w_1$, \ldots, $w_{n-1}$. For every $i \in \{0, 1, \ldots, n-1 \}$, we attribute to the word $w_i$ an `energy level' equal to the ranking of $w_i$ in ${\cal T}$, that is, we set, for every $i \in \{0, 1, \ldots, n-1 \}$, $E_i=i$.\footnote{Unlike physics where energy has a (derived) unit of measurement, energy is expressed in dimensionless units in our theoretical framework for language and cognition. We also mention that setting $E_0=0$ is the typical choice that is made in physics when dealing with statistical behaviour at low temperatures. But, other choices are likewise possible, as what is really important is the difference between energy levels.} Thus, the most frequent word is attributed the lowest energy level, with $E_0=0$ being the `ground energy level', and we have $n$ different energy levels. Then, for every $i \in \{ 0, 1, \ldots, n-1 \}$, let $N(E_i)$ be the number of appearance of the word $w_i$ in ${\cal T}$, so that the `total number of the words' in the text ${\cal T}$ is 
\begin{equation} \label{totalnumber}
N=\sum_{i=0}^{n-1} N(E_i)
\end{equation}

For every $i \in \{ 0, 1, \ldots, n-1 \}$, each single word $w_i$ is associated with a conceptual entity, which we call `cogniton', in a `single-entity state', or `micro-state' $p_i$, thus there are $N(E_i)$ entities in the micro-state $p_i$. For every word $w_i$, the $N(E_i)$ entities are `identical' and have a defined energy $E_i=i$, hence the `total energy' of the word $w_i$ is $E_i N(E_i)=i N(E_i)$, $i \in \{ 0, 1, \ldots, n-1 \}$. For every $i,j \in \{ 0, 1, \ldots, n-1 \}$, $i \ne j$, the words $w_i$ and $w_j$ are instead `non-identical'. Indeed, while two of the $N(E_i)$ words $w_i$ can be exchanged in the text ${\cal T}$ without altering its overall meaning, an exchange of two different words $w_i$ and $w_j$, $i\ne j$, will generally alter the overall meaning of ${\cal T}$. 

To sum up, a text ${\cal T}$ is an entity of meaning composed by $N$ cognitons, as expressed by Eq. (\ref{totalnumber}), whose `total energy' is 
\begin{equation} \label{totalenergy}
E=\sum_{i=0}^{n-1} E_i N(E_i)=\sum_{i=0}^{n-1} i N(E_i)
\end{equation}

If ${\cal T}$ is a text of a medium- or large-size, as those that can be found in linguistic or literary corpora of documents, it is reasonable to assume that we are in the typical conditions that are assumed in (classical and quantum) statistical mechanics, namely, ${\cal T}$ can be considered as an `ideal gas of cognitons in thermodynamic equilibrium' \citep{aertsbeltran2020,aertsbeltran2022a,aertsbeltran2022b,independence2024}. Moreover, if we choose a given text ${\cal T}$ from one of these corpora, its total number of words $N$ and total energy $E$ in Eqs. (\ref{totalnumber}) and (\ref{totalenergy}), respectively, can be easily calculated and are constants that characterise the text in question.

Then, regarding the statistical distribution to be satisfied by the numbers of appearance $N(E_i)$, $i \in \{ 0, 1, \ldots, n-1\}$, if we come back to our considerations in Sect.~\ref{intro}, we have that the `Maxwell--Boltzmann distribution'
\begin{equation} \label{mb}
N_{MB}(E_i)=\frac{1}{Ce^{\frac{E_i}{D}}}
\end{equation}
is typically derived in classical statistical mechanics under the assumption that the $N$ entities are `identical and distinguishable' and described by classical mechanics. In this case, the different micro-states are statistically independent, that is, the probability that one of the $N$ entities is in a given micro-state does not affect the probability that another of the $N$ entities is in the same micro-state (see also footnote \ref{statind}). 

On the other side, in quantum statistical mechanics, the requirement of complete indistinguishability of identical entities described by QM, incorporated in the symmetry exchange postulate (see Sect.~\ref{intro}), entails that either the `Fermi--Dirac distribution'
\begin{equation} \label{fd}
N_{FD}(E_i)=\frac{1}{Ae^{\frac{E_i}{B}}+1}
\end{equation}
or the `Bose--Einstein distribution' 
\begin{equation} \label{be}
N_{BE}(E_i)=\frac{1}{Ae^{\frac{E_i}{B}}-1}
\end{equation}
should be satisfied.\footnote{We will not deal with the Fermi--Dirac distribution in the present article, as it does not typically occur in cognitive phenomena connected with human language, at least, not at the level of analysis proposed here. On the contrary, it is likely that this type of statistical distribution occurs naturally in human memory (see, e.g., \citet{aertsbeltran2022a,aertsbeltran2022b}).} More precisely, fermions, e.g., electrons, obey the Fermi--Dirac distribution, whereas bosons, e.g., photons, obey the Bose--Einstein distribution. As anticipated in Sect.~\ref{intro}, in both cases, identical quantum entities are statistically dependent, more precisely, electrons cannot occupy the same micro-state, whereas bosons do not have this limitation and rather tend to clump, or cluster, together in the same micro-states \citep{eisbergresnick1985,huang1987}. In the latter case, an extreme situation is obtained if one lowers the temperature of an ideal gas till a point when all bosons put themselves in the same state, which is the lowest energy state, and the gas behaves as a Bose--Einstein condensate \citep{cornellwieman2002,ketterle2002,annett2005}. 

The parameters $A$ and $B$ (similarly, $C$ and $D$) can be easily determined in physics. Indeed, one gets
\begin{equation}
B=kT
\end{equation}
where $k$ is `Boltzmann's constant' and $T$ is the `absolute temperature', and
\begin{equation} \label{fugacity}
A=e^{-\frac{\mu}{kT}}
\end{equation}
where $\mu$ is the `chemical potential' of the gas. One typically introduces another parameter, $f=e^{\frac{\mu}{KT}}$, called the `fugacity' of the gas \citep{huang1987,eisbergresnick1985}. The inverse formulas $f=\frac{1}{A}$ and $\mu=-B \ln A$ easily follow too. 

Now, to experimentally detect the best statistical distribution of the words in a text, let us choose a text within a given corpus of documents. Since the numbers $N$ and $E$ in Eqs. (\ref{totalnumber}) and (\ref{totalenergy}), respectively, can be calculated from the empirical numbers of appearance $N(E_i)$, $i \in \{ 0, 1, \ldots, n-1 \}$, as anticipated above, one can calculate the pairs $(C,D)$ and $(A,B)$, hence $N_{MB}(E_i)$ and $N_{BE}(E_i)$, in Equations (\ref{mb}) and (\ref{be}), respectively, such that
\begin{eqnarray}
&&\sum_{i=0}^{n-1} \frac{1}{Ce^{\frac{E_i}{D}}}= N \\
&&\sum_{i=0}^{n-1} \frac{E_i}{Ce^{\frac{E_i}{D}}}= E \\
&&\sum_{i=0}^{n-1} \frac{1}{Ae^{\frac{E_i}{B}}-1}= N \\
&&\sum_{i=0}^{n-1} \frac{E_i}{Ae^{\frac{E_i}{B}}-1}= E
\end{eqnarray}  

We have recently  studied a medium-size text, the story ``In Which Piglet Meets a Heffalump'' taken from Milne's ``Winnie the Pooh'' (1926). The results are summarized in Tab. \ref{table1}. We do not report complete details of the analysis, but limit ourselves to sketch the results here (the interested reader can make reference to \citet{aertsbeltran2020,aertsbeltran2022a,aertsbeltran2022b,beltran2021,philtransa2023,independence2024}). The Bose--Einstein distribution provides an almost perfect fit of the experimental numbers of appearance $N(E_i)$, while the Maxwell--Boltzmann distribution does not provide a good fit at all. Hence, the ``Winnie the Pooh'' story statistically behaves as an ideal gas of bosons, which justifies the term ``cogniton'', meant as the fundamental quantum of human language. We have also recently studied two large-size texts, namely, Well's ``The Magic Shop'' (1903) and Swift's ``Gulliver's Travels'' (1726), obtaining exactly the same results, which are also reported in Tab. \ref{table1}.
 
\begin{footnotesize}
\begin{table}
\begin{center}
\begin{tabular}{ |c|c|c|c|c|c|c|c| } 
\hline 
Text & Levels & Tot. words & Tot. energy & \multicolumn{4}{|c|}{Distribution} \\   
  &  &  & & \multicolumn{2}{|c}{BE} & \multicolumn{2}{c|}{MB} \\ 
\hline 
${\cal T}$ & $n$ & $N$ & $E$ & $A$ & $B$ & $C$ & $D$ \\
\hline
WtP & 543 & 2,655 & 242,891 & 1.0078 & 593.51 & 0.0353 & 93.63 \\
TMS & 3,500 & 3,934 & 817,415 & 1.0005 & 722.05 & 0.0531 & 108.28 \\
GT & 8,294 & 103,184 & 68,903,521 & 1.0002 & 19,356.22 & 0.0075 &1,355.31 \\ 
\hline 
\end{tabular}
\caption{We report the number of different words, i.e. the number of energy levels, $n$, the total number of words $N$ and the total energy of words $E$ for three story-telling texts written in English language: Winnie the Pooh (WtP), The Magic Shop (TMS) and Gulliver's Travels (GT). 
We also indicate the value of the constants $C$ and $D$ of the Maxwell-Boltzmann distribution in Eq. (\ref{mb}) and the value of the constants $A$ and $B$ of the Bose--Einstein distribution in Eq. (\ref{be}) that fit empirical data. \label{table1}}
\end{center}
\end{table}
\end{footnotesize}

The results sketched above significantly show that Bose--Einstein statistics is the governing statistics for the word distribution in a story-telling text, which can be considered as an ideal Bose--Einstein gas of identical cognitons. Moreover, let $w_i$ and $w_j$, with $i<j$, $i,j\in \{0, 1, \ldots, n-1 \}$, be two different words of a given text ${\cal T}$, and suppose that additional paragraphs are added to ${\cal T}$. It is then reasonable to expect that the more frequent word $w_i$ has a higher probability than $w_j$ to occur in the new paragraphs, as the words that are added have to match the overall meaning of the text ${\cal T}$. That is, the different words, that make up the story and correspond to different micro-states of the cogniton, lack statistical independence. This is the linguistic-conceptual counterpart of the typical tendency of bosons to clump, or cluster, together in the same micro-states. We also notice that cognitons, however, are not completely indistinguishable, because identical words are indistinguishable, but non-identical words are not.

The three aspects above, namely, presence of Bose--Einstein statistics, lack of statistical independence and incomplete indistiguishability, that we have identified in texts of English language, will be found back and widely extended when dealing with texts of Italian language, as we intend to prove in Sect.~\ref{experiment}.\footnote{Though not the subject of the present article, we remind that the quantum theoretical framework illustrated in this section also provides a theoretical foundation of `Zipf's law' \citep{zipf1935,zipf1949}, which is systematically present in data across a range of cultural areas, even outside linguistic domains, and whose theoretical origin is currently unknown. Zipf's law holds when Italian language is used too \citep{tuzzipopescualtmann2009}.}

\section{An empirical study through literary corpora of Italian language\label{experiment}}
We present in this section the results and analysis of our empirical study on various texts of Italian literature selected from a corpus of documents of Italian language. As we will see, the results presented in this section confirm, strengthen and extend those sketched in Sect.~\ref{theory} on texts produced in English language.

In our empirical study, we selected five literary texts of Italian authors of XIX and XX century within the freely available Italian digital library ``BibIt'', see the webpage \url{http://www.bibliotecaitaliana.it}. More precisely, we have chosen ``Del Romanzo Storico'' (Alessandro Manzoni, 1830), ``Cuore'' (Edmondo De Amicis, 1886), ``Le Avventure di Pinocchio'' (Carlo Collodi, 1883), ``Senilità'' (Italo Svevo, 1898), and ``Le Ultime Lettere di Jacopo Ortis'' (Ugo Foscolo, 1802). While, `all' studies showed the same empirical pattern with respect to their statistical behaviour, we will concentrate on the novels ``Cuore'' (English translation ``Heart'') and ``Del Romanzo Storico'' (English translation ``On the Historical Novel'') for our considerations in this section, for the sake of brevity. The empirical results of all texts are reported in Tab. \ref{table2}.

\begin{footnotesize}
\begin{table}
\begin{center}
\begin{tabular}{ |c|c|c|c|c|c|c|c| } 
\hline 
Text & Levels & Tot.words & Tot. energy & \multicolumn{4}{|c|}{Distribution} \\   
  &  &  & & \multicolumn{2}{|c}{BE} & \multicolumn{2}{c|}{MB} \\ 
\hline 
${\cal T}$ & $n$ & $N$ & $E$ & $A$ & $B$ & $C$ & $D$ \\
\hline
RS & 5,005 & 22,600 & 7,706,035 & 1.0014 & 2,050.39 & 0.0273 & 309.68 \\
Cu & 10,483 & 77,826 & 83,981,879 & 1.0004 & 10,372.82 & 0.0139 & 1,080.24 \\
Pi & 6,108 & 38,436 & 28,022,286 & 1.0011 & 5,988.58 & 0.0190 & 731.00 \\
Se & 9,024 & 63,339 & 60,296,678 & 1.0005 & 8,713.51 & 0.0151 & 953.17 \\
JO & 9,607 & 44,571 & 25,544,173 & 1.0008 & 3,562.03 & 0.0245 & 518.82 \\
\hline 
\end{tabular}
\caption{We report the number of different words, i.e. the number of energy levels, $n$, the total number of words $N$ and the total energy of words $E$ 
 for five story-telling texts written in Italian language: Romanzo Storico (RS), Cuore (Cu), Pinocchio (Pi), Senilit\`a (Se) and Jacopo Ortis (JO). 
We also indicate the value of the constants $C$ and $D$ of the Maxwell-Boltzmann distribution in Eq. (\ref{mb}), and the value of the constants $A$ and $B$ of the Bose--Einstein distribution in Eq. (\ref{be}), that fit empirical data. \label{table2}}
\end{center}
\end{table}
\end{footnotesize}

Let us apply the quantum theoretical framework elaborated in Sect.~\ref{theory} to the text ``Cuore'' (``Heart''). The most frequent word was \emph{Di}, which was then attributed an energy level $E_0=0$, with a number of appearance $N(E_0)=2,180$, followed by the word \emph{Che}, which was attributed an energy level $E_1=1$, with a number of appearance $N(E_1)=2,071$. Going down, the 22nd ranked word \emph{Father} was attributed an energy level $E_{21}=21$, with a number of appearance $N(E_{21})=426$. Hence, the total energy of the word \emph{Father} was $E_{21} N(E_{21})=21 \times 426=8,946$, and so on.

Next, to determine the statistical distribution for the numbers of appearance $N(E_i)$ that best fitted empirical data, whether the Maxwell--Boltzmann $N_{MB}(E_i)$ or the Bose--Einstein $N_{BE}(E_i)$ distribution, we computed the values of the constants $C$ and $D$ in Eq. (\ref{mb}) such that
\begin{eqnarray}
&&\sum_{i=0}^{n-1} \frac{1}{Ce^{\frac{E_i}{D}}}= 82,603 \\
&&\sum_{i=0}^{n-1} \frac{E_i}{Ce^{\frac{E_i}{D}}}= 89,055,352
\end{eqnarray}  
and the values of the constants $A$ and $B$ in Eq. (\ref{be}) such that
\begin{eqnarray}
&&\sum_{i=0}^{n-1} \frac{1}{Ae^{\frac{E_i}{B}}-1}= 82,603 \\
&&\sum_{i=0}^{n-1} \frac{E_i}{Ae^{\frac{E_i}{B}}-1}= 89,055,352
\end{eqnarray}

\begin{figure}
\centering
\includegraphics[width=9cm]{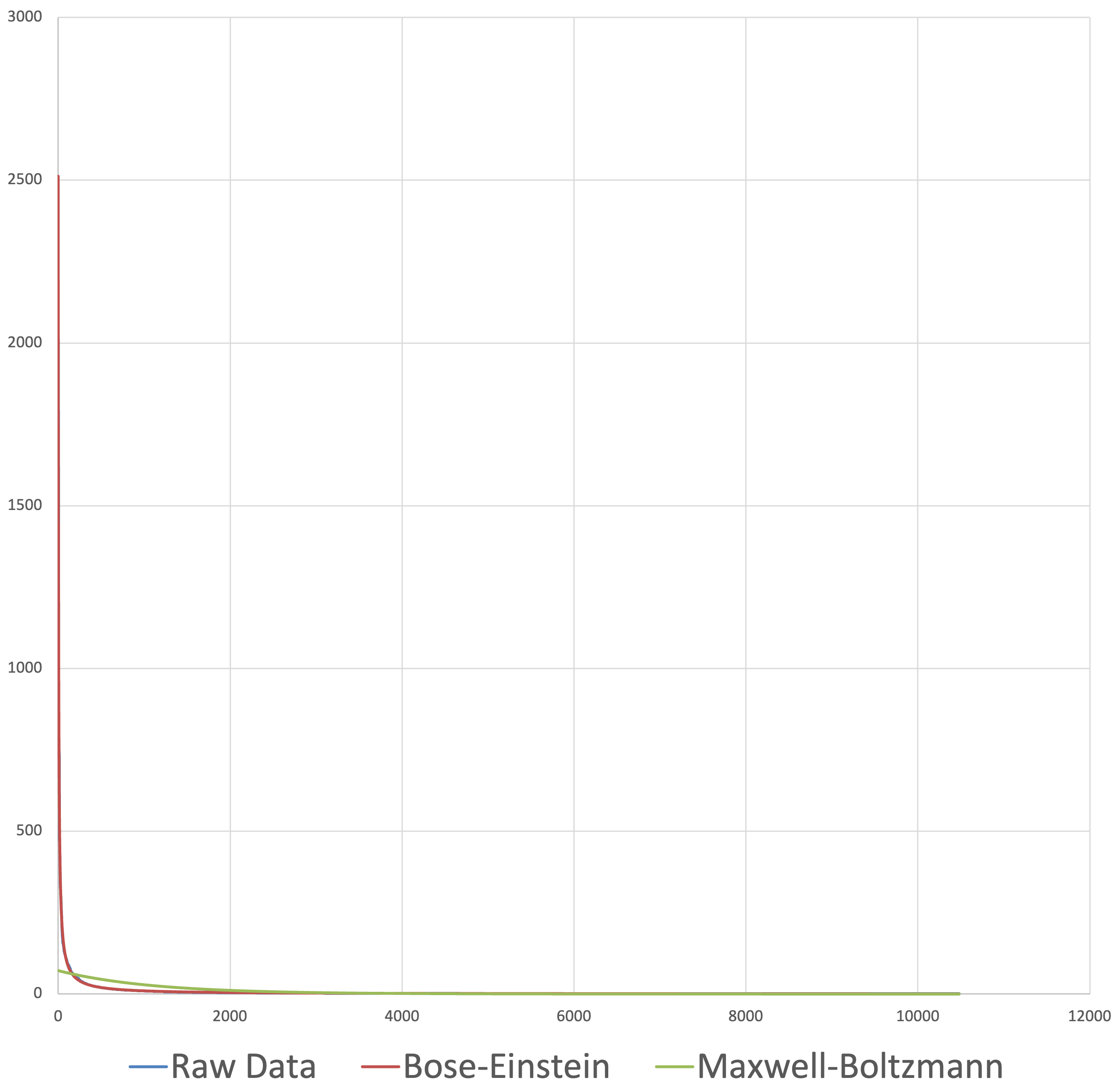}
\caption{We report the numbers of appearance $N(E_i)$ of the words in the text ``Cuore'' (``Heart''), ranked from lowest energy level, i.e. the most frequent word, to highest energy level, i.e. the least frequent word. The blue graph (almost entirely below the red graph, see Fig.~\ref{heart_2})  
corresponds to empirical data, i.e. the collected numbers of appearance from the text, the red graph is a Bose--Einstein distribution model for the same numbers of appearance, and the green graph  is a Maxwell--Boltzmann distribution model. The red and blue graphs coincide almost completely, whereas the green graph presents large deviations from the blue graph of the data. This shows that the Bose--Einstein distribution does provide a good model for the numbers of appearance, while the Maxwell--Boltzmann distribution does not.}%
    \label{heart_1}%
\end{figure}

\begin{figure}
\centering
\includegraphics[width=9cm]{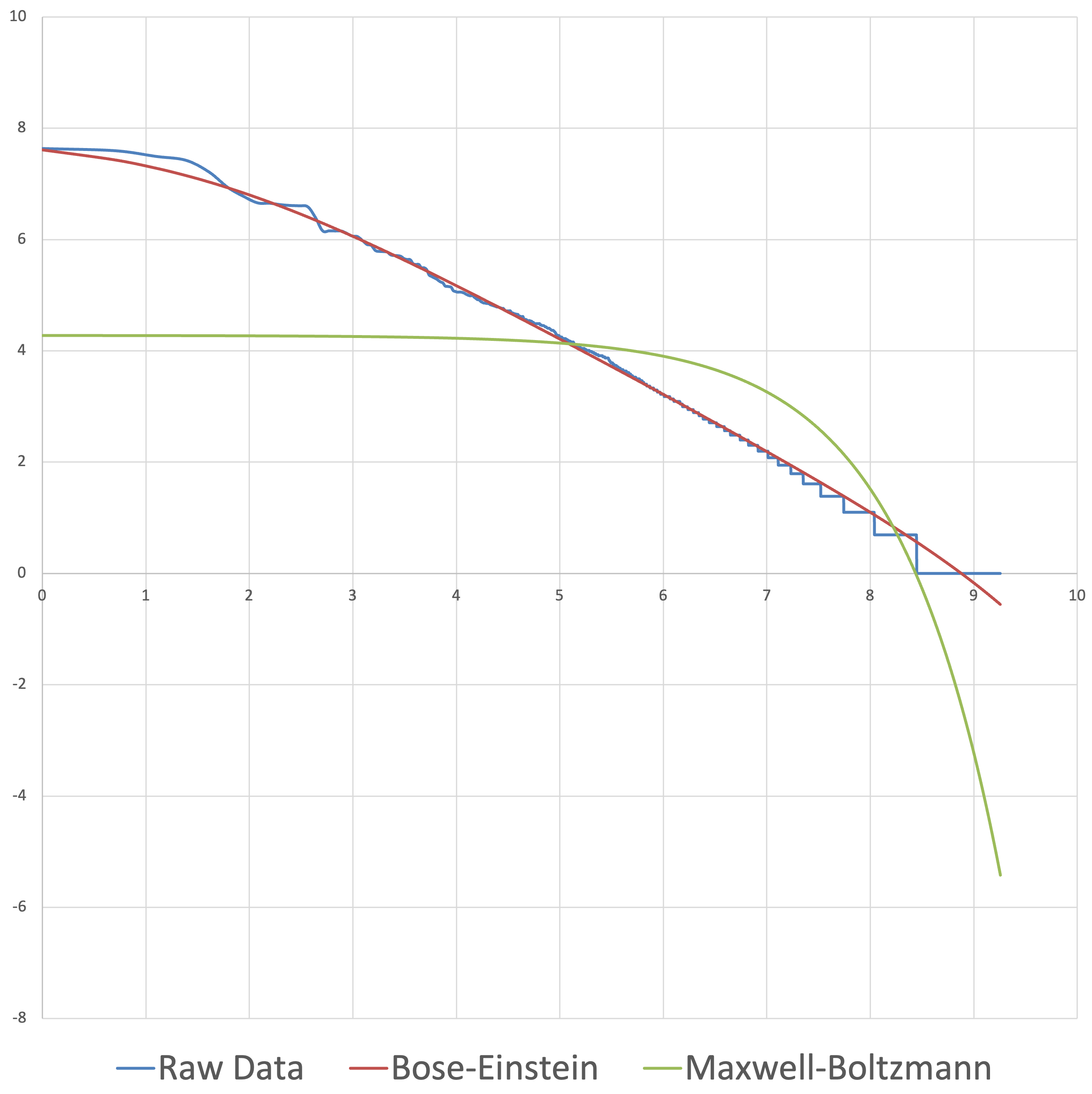}
    \caption{We report the $\log / \log$ graphs of the numbers of appearance of Fig.~\ref{heart_1}, and their Bose--Einstein and Maxwell--Boltzmann distribution models. As already noted in Fig.~\ref{heart_1}, the red and blue graphs coincide almost completely, whereas the green graph presents large deviations from the blue graph of the data. This shows that the Bose--Einstein distribution does provide a good model for the numbers of appearance, while the Maxwell--Boltzmann distribution does not.}%
    \label{heart_2}%
\end{figure}

\begin{figure}
\centering
\includegraphics[width=9cm]{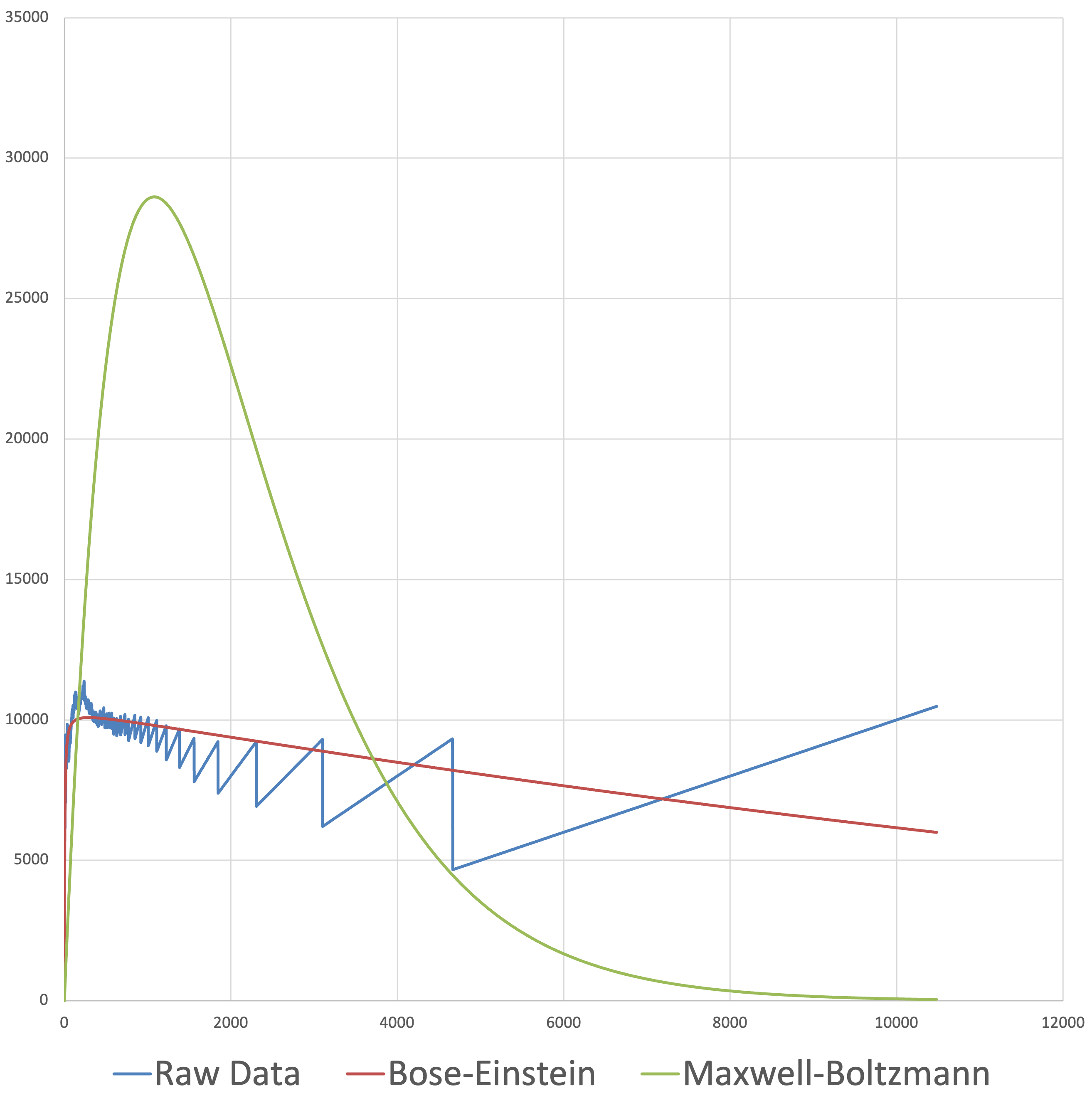}
\caption{We report the energy distribution of the text ``Cuore'' (``Heart''). More precisely, the blue graph reports the energy $E_iN(E_i)$ of the text per energy level $E_i=i$, the red graph reports the same energy per energy level as modelled by the Bose--Einstein distribution, while the green graph, reports the energy per energy level as modelled by the Maxwell--Boltzmann distribution.}
\label{heart_3}
\end{figure}

The results are presented in Fig.~\ref{heart_1}, where the empirical numbers of appearance $N(E_i)$ are compared with the numbers of appearance predicted by the Bose--Einstein distribution model $N_{BE}(E_i)$ and the numbers of appearance predicted by the Maxwell--Boltzmann distribution model $N_{MB}(E_i)$, and in Fig.~\ref{heart_2}, where the three corresponding $\log / \log$ distributions are also compared. Figure~\ref{heart_3}, instead,  
reports the total energies $E_i N(E_i)$ per energy level $E_i=i$ of the empirical data, again compared to those predicted by a Bose--Einstein and Maxwell--Boltzmann distribution models.

As we can see, in both Figs.~\ref{heart_1} and \ref{heart_2}, the Bose--Einstein distribution provides a very good theoretical model of the empirical data, whereas large deviations are observed from the Maxwell--Boltzmann distribution.

We applied the quantum theoretical framework to the word distribution of all five texts, always finding that the Bose--Einstein distribution models well the collected data, whereas the Maxwell--Boltzmann distribution does not provide a good model for them. These results confirm and strengthen those obtained on texts of English language (see Sect.~\ref{theory}) \citep{aertsbeltran2020,aertsbeltran2022a,aertsbeltran2022b}, extending them to texts produced in Italian language. The conclusion which directly follows is that, as in the English language case, Bose--Einstein statistics represents the statistical behaviour of the words that are contained in a text produced by Italian language. This suggests that `there is an intrinsic cognitive dynamics which is independent of the specific language that is used and governs the formation and combination of the concepts corresponding to the words that appear in a text produced by human language'. In this dynamics, meaning plays a fundamental role, as we will see in Sects. \ref{randomization} and \ref{explanation}. But, we can already at this stage make some relevant considerations on the appearance of a boson-type quantum statistics, as follows.  

Firstly, we known from Sect.~\ref{intro} that the appearance of a quantum statistical behaviour is attributed to the complete indistinguishability of identical entities in quantum mechanics, which is formally incorporated in the symmetry exchange postulate. However, some authors, including ourselves, maintain that this postulate is problematical (see also Sect.~\ref{physics}). On the other side, we have seen in Sect.~\ref{theory} that identical words in a text are indistinguishable, while different words are not indistinguishable. We remind that identical words correspond to the same energy level, hence the same micro-state, of the cogniton, while different words generally correspond to different energy levels, hence different micro-states, of the cogniton. 
Let us take the example of the the word \emph{Romanzo} (meaning \emph{Novel} in English) in the text ``Del Romanzo Storico'' (``On the Historical Novel''), which has an energy level $E_{51}=51$ and a number of appearance $N(E_{51})=52$. 
It is evident that if we take two places where the word \emph{Romanzo} appears, and exchange the two words, the overall meaning of the text does not change. 
This is because \emph{Romanzo}, as it emerges as a concept in the meaning that the story holds, is an abstract entity within the story, i.e. a concept and not an object, and therefore identical and indistinguishable from any other word \emph{Romanzo} that appears in the story. We believe that the appearance of quantum structures in large texts is partly due to the fact that a text that tells a story is primarily an entity of meaning, hence the words in it are conceptual and not objectual. 
Secondly, let us consider the word \emph{Storia} (meaning \emph{History} in English) which has an energy level $E_{18}=18$ and a number of appearance $N(E_{18})=139$ in the text ``Del Romanzo Storico'', and consider again the word \emph{Romanzo}. 
It is evident that the two words \emph{Romanzo} and \emph{Storia}, hence the corresponding micro-states, are not statistically independent, in the sense specified in Sects. \ref{intro} and \ref{theory}. 
Indeed, suppose that some additional paragraphs are now added to the text. Since \emph{Storia} occurs $139$ times, while \emph{Romanzo} occurs $52$ times, we expect that the probability that \emph{Storia} occurs in the additional paragraphs is higher than the probability that \emph{Romanzo} occurs in them. This is because the entire text carries an overall meaning, hence the words whose meaning is closer to that overall meaning have a higher chance to appear if the text is continued to be written. 
Using the terminology of physics, the more cognitons are in a given micro-state, the more a new cogniton is likely to be in that micro-state. Equivalently, the more a word is used in a text, the more likely it is that the same word will appear in a new piece of text based on the previous one. This expresses the typical tendency of bosons to clump, or cluster, together in the same micro-states. We believe that this lack of statistical independence in non-identical, hence not indistinguishable, words, is partly responsible for the appearance of Bose--Einstein statistics in the statistical behaviour of large texts.

The two considerations above again confirm and strengthen similar considerations we made for texts produced in the English language. Moreover, they clearly indicate that meaning plays a fundamental role in the conceptual combination of the words in a text, because any text has an `overall coherence' in terms of meaning, hence each word that is added to the text has to be `connected' with all the words that are already present in the text in such a way that this meaning coherence is preserved. However, before we can fully grasp the role that meaning might play in the linguistic-conceptual domain, we need to report the results of a variant of the empirical study presented in this section. This is the aim of Sect.~\ref{randomization}.

\section{The effect of word randomization\label{randomization}}
We present in this section the results of an effect of `word randomization' which we have firstly introduced in \citet{independence2024} and may help clarifying how meaning relates to the appearance of quantum statistics in language.

Let us consider again the five literary texts of Italian language analysed in Sect.~\ref{experiment} and investigate how the introduction of `randomness' may affect the statistical distribution of the words appearing in the texts. The underlying idea is to partially remove the meaning of a text and check whether and how this removal will impact the statistics distribution of its words. Let us focus on the text ``Senilità'' (``As a Man Grows Older''), for the sake of brevity. We stress, however, that the results obtained on this text generally hold for all the others too. 
The text ``Senilità'' contains $n=9,024$ different words, hence $9,024$ energy levels, and a total number of words $N=63,339$ (see Tab. \ref{table2}). Let us introduce the notion of `placeholder', which will simplify our considerations. 
In the text ``Senilità'', which we will call the `original text' from now on, each word is in a given placeholder, thus we have $N=63,082$ placeholders overall. But, it is clear that a given word of the original text appears in various different placeholders, depending on its number of appearance in that text.

We have used the tools offered by the website \url{www.random.org} to change the relative frequencies with which the different words of the ``Senilità'' text occupy the available $63,339$ placeholders, determining new numbers of appearance in a  genuinely random way.\footnote{We note that the method of randomization developed in the website is linked to weather and atmospheric noise, thus it involves gases at room temperature. This is relevant to our scopes, because we will see in this and the following sections that word randomization can be considered as the linguistic analogue of a temperature increase. \label{weather}} This operation of word randomization generates a new text with the same number $n=9,024$ of different words, hence again $9,024$ energy levels, and the same total number of words $N=63,339$, as the original text. We will call `randomized text' from now on the new version of ``Senilità'' generated in this way. It should be noticed that this operation of word randomization is a subtle reshuffling of the relative frequencies of appearance of words, as it will become clear in the following. 

Let us preliminary state that this process of word randomization has  a profound impact on the energy level carried by each word. Indeed, in the randomized text, the most frequent word has a number of appearance equal to $21$, the second most frequent word has a number of appearance equal to $19$, and so on.
Let us now apply the quantum theoretical framework in Sect.~\ref{theory} and assign again energy levels according to the decreasing number of appearance. In this way, we have generated a randomized text that has the same energy levels $E_0$, $E_1$, \ldots, $E_{n-1}$ as the original text but different numbers of appearance $N(E_0)$, $N(E_1)$, \ldots, $N(E_{n-1})$, i.e. different occupation numbers of these energy levels, which now correspond to a lower occupation of the lower energy levels.\footnote{More specifically, regarding the text ``Senilità'', we generated through the above mentioned website an amount $N=63,339$ of numbers taken out of the $n=9,024$ numbers. We made different trials and found that, while the most frequent number was different in each trial, the numbers of appearance were reasonably stable across trials. This operation was repeated on all texts discussed in Sect.~\ref{experiment}.} It is also important to notice that we have not entirely removed the meaning of the original text, because each word appearing in the randomized text has its own meaning and the randomized text still contains the same words and with the same total number as the original text. However, the numbers of appearance of each of the different words in the original text are now randomly changed, 
which entails that we have removed some of the meaning generated when the original text was written. We believe that the procedure described above has far reaching consequences on the development of a thermodynamic theory of language. Some of these consequences will be examined in detail in a forthcoming article \citep{temperature2025}. But, we can already at this stage draw some interesting conclusions, as follows.

In its original (non-randomized) version, ``Senilità'' showed the statistical behaviour illustrated in Figs. \ref{heart_1}-\ref{heart_3}, hence the words in it obey Bose--Einstein statistics and show large deviations from Maxwell--Boltzmann statistics. We may pthen wonder whether the statistical distribution of the words contained in the randomized version can still be modelled by the Bose--Einstein distribution and, more important, whether the Maxwell--Boltzmann distribution still does not provide a good model at all. 
To this end, we apply the quantum theoretical framework presented in Sect.~\ref{theory} to the randomized version of the text ``Senilità'' and determine the statistical distribution for the numbers of appearance $N(E_i)$ that best fit the new empirical data, computing the pairs $(C,D)$ in Eq. (\ref{be}) and $(A,B)$ of Eq. (\ref{be}) subject to the conditions that the total number of words has to be $N=63,339$, as in the original text, while their total energy has now risen to $E=225,613,236$ (it was $E=60,296,678$ in the original text).

\begin{figure}
\centering
\includegraphics[width=8cm]{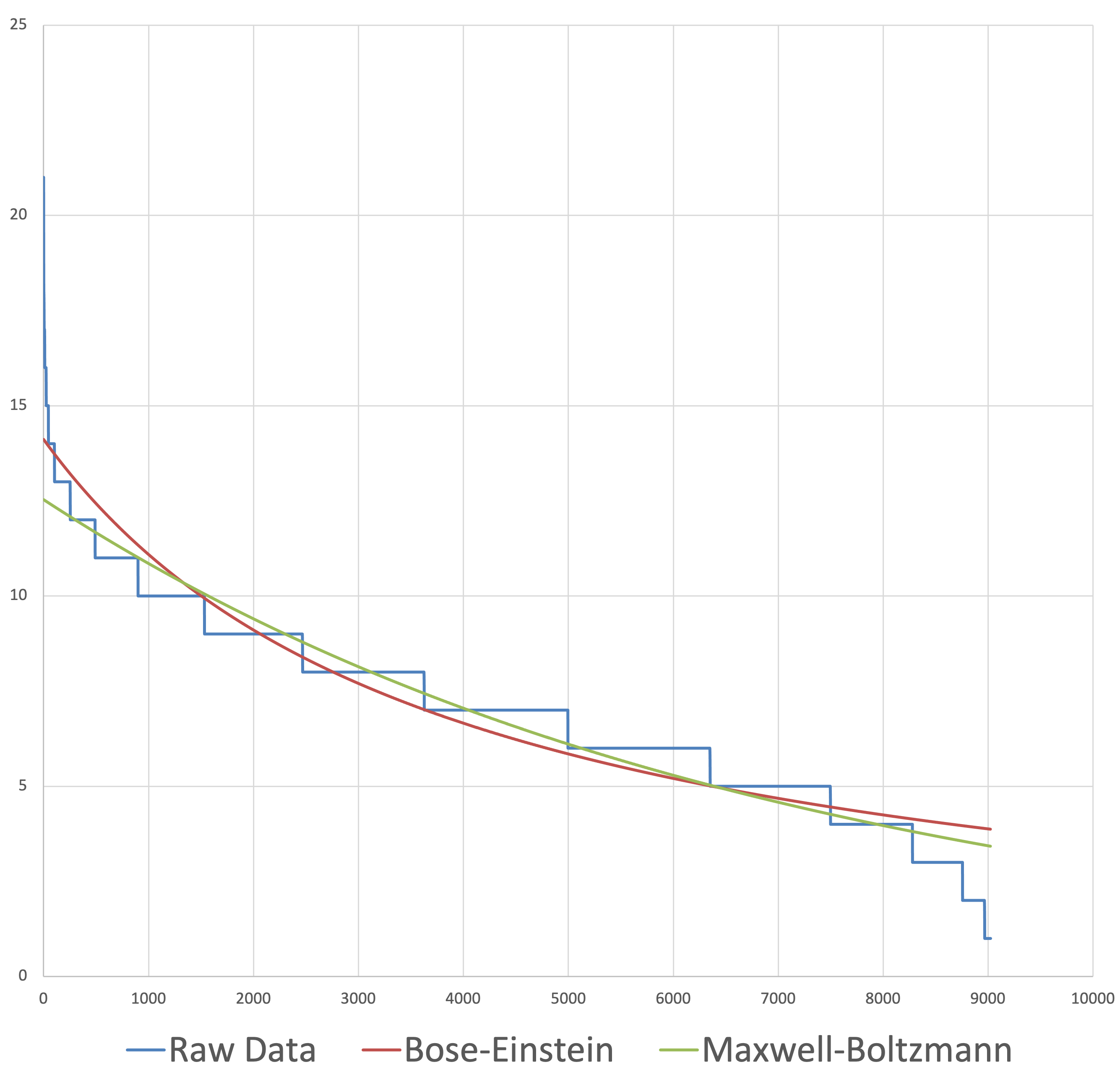}
\caption{We report the numbers of appearance $N(E_i)$ of the words in the randomized version of the text ``Senilità'' (``As a Man Grows Older''), ranked from lowest energy level, corresponding to the most frequent word, to highest energy level, corresponding to the least frequent word. The blue graph represents empirical data, i.e. the collected numbers of appearance from the randomized text, the red graph is a Bose--Einstein distribution model for the same numbers of appearance, and the green graph is a Maxwell--Boltzmann distribution model. The Bose--Einstein distribution is still a good model for the numbers of appearance, but the quality of the Maxwell--Boltzmann distribution model is now much better.}%
    \label{senilitarandom_1}%
\end{figure}

\begin{figure}
\centering
\includegraphics[width=9cm]{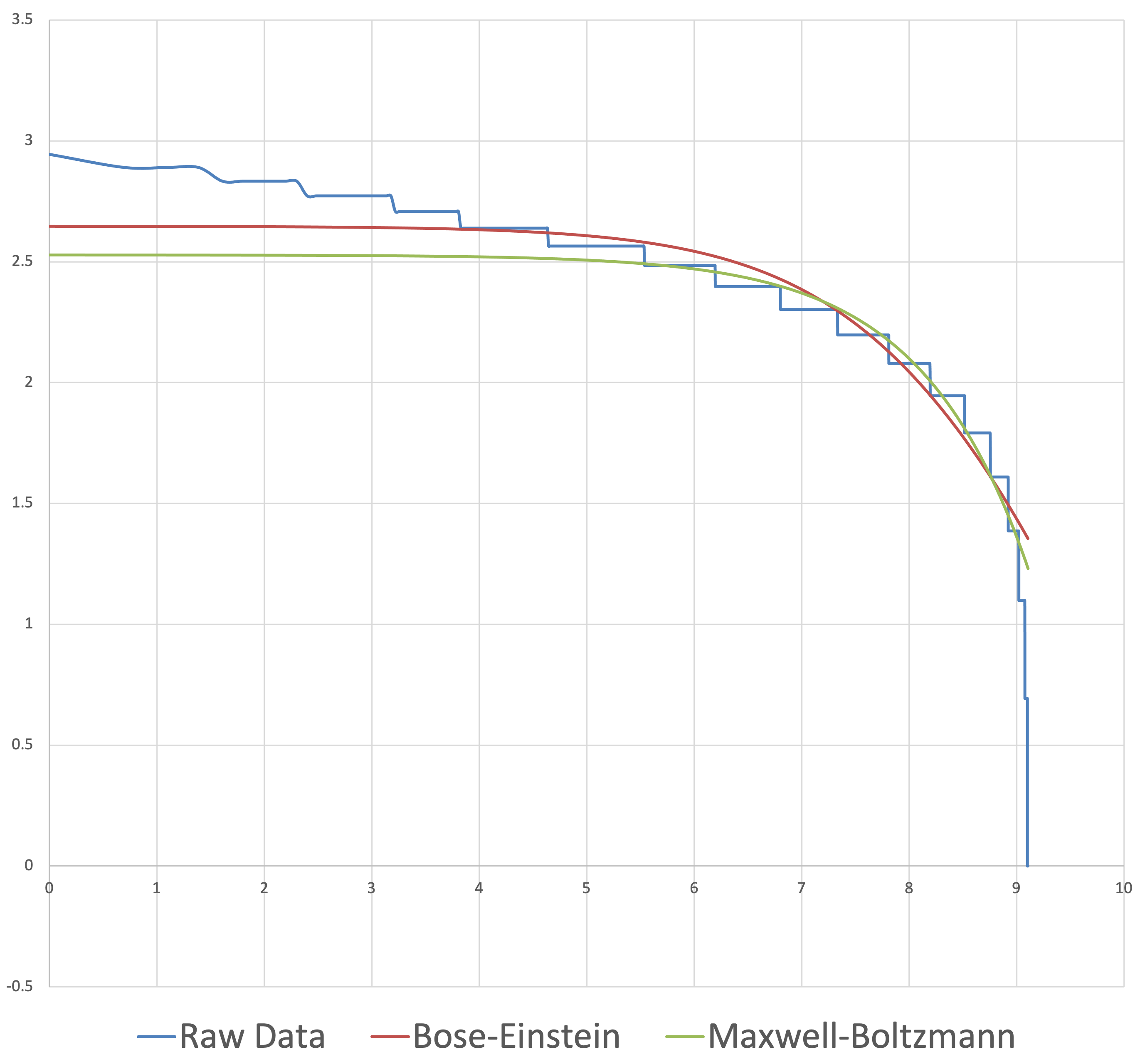}
\caption{We report the $\log / \log$ graphs of the numbers of appearance of Fig.~\ref{senilitarandom_1}, and their Bose--Einstein and Maxwell--Boltzmann distribution models. As already noted in Fig.~\ref{senilitarandom_1}, the Bose--Einstein distribution is still a good model for the numbers of appearance, but the quality of the Maxwell--Boltzmann distribution model is now much better.}%
    \label{senilitarandom_2}%
\end{figure}

\begin{figure}
\centering
\includegraphics[width=9cm]{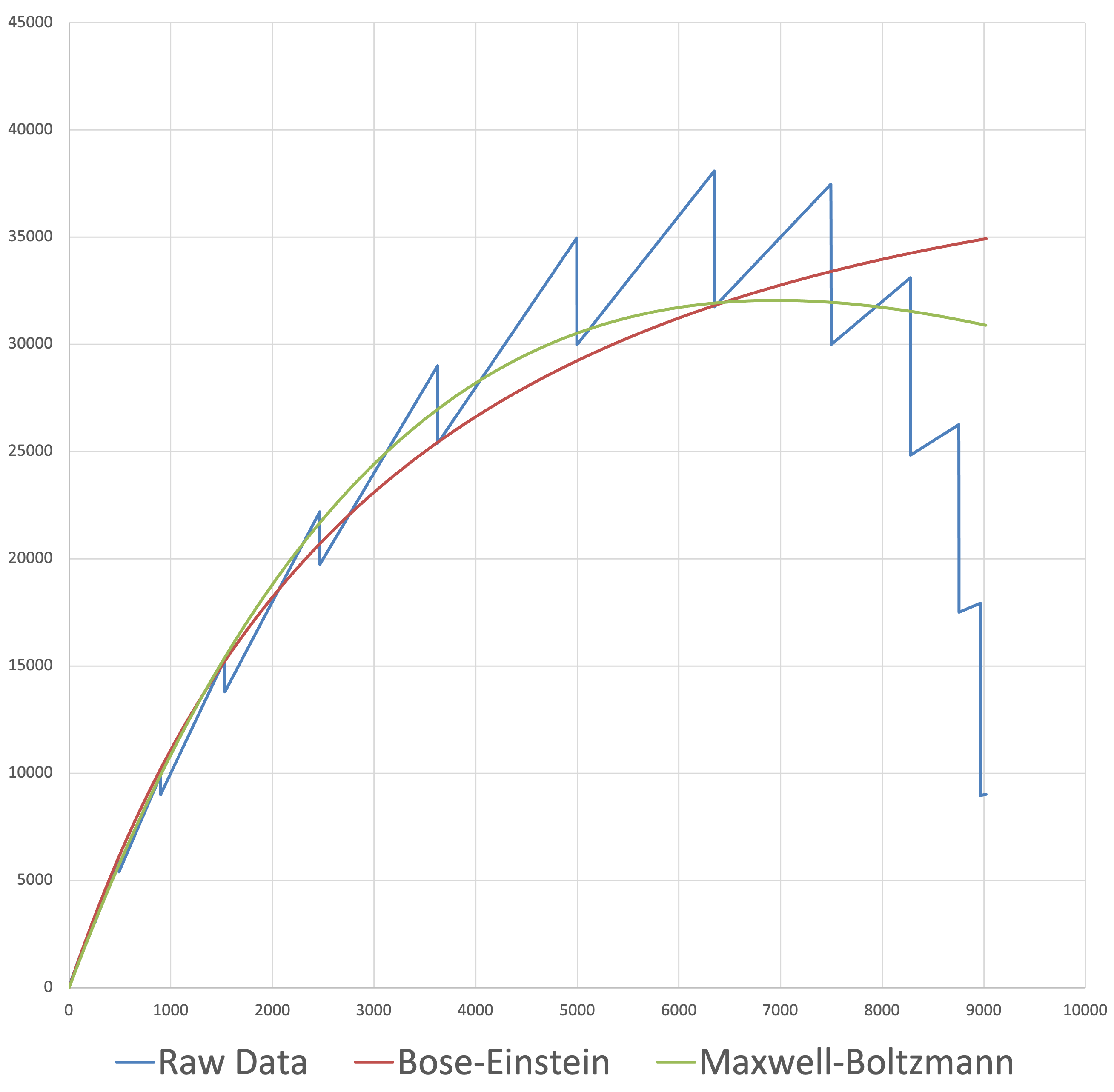}
\caption{We report the energy distribution of the randomized version of the text ``Senilità'' (``As a Man Grows Older''). The blue graph reports the energy $E_iN(E_i)$ of the story text per energy level $E_i=i$, the red graph reports the same energy per energy level as modelled by the Bose--Einstein distribution, and same occurs for the green graph, which follows the Maxwell--Boltzmann distribution.}
\label{senilitarandom_3}
\end{figure}

The results are presented in Fig.~\ref{senilitarandom_1}, where the empirical numbers of appearance $N(E_i)$ are again compared with the numbers of appearance predicted by the Bose--Einstein and Maxwell-Boltzmann distributions, and in Fig.~\ref{senilitarandom_2}, where the corresponding $\log / \log$ distributions are compared too. Finally, Figure \ref{senilitarandom_3} reports the overall energies per energy level of the empirical data of the randomized text, compared to those predicted by the Bose--Einstein and Maxwell--Boltzmann distributions.

As we can see by looking at Figures~\ref{senilitarandom_1}-\ref{senilitarandom_3}, the Bose--Einstein distribution still provides a reasonably good model for the empirical data of the randomized text, but the Maxwell--Boltzmann distribution has now much improved as a model of the data. Indeed, the difference between the two distribution models is now much smaller than in the original text, with the Bose--Einstein distribution working better at lower energies and the Maxwell--Boltzmann distribution working better at higher energies.\footnote{We observe that when the temperature of a system increases, the available thermal energy is much greater than the separation between the energy levels of the microstates. This means that the entities forming the system have enough energy to occupy many different microstates. Under these conditions, it is natural to expect quantum correlations to become less relevant, making the different microstates more independent of each other, which is a characteristic of the Maxwell--Boltzmann distribution.}

To understand the result above, let us consider Tab.~ \ref{table3}, where we report the total energy of words $E$, together with the fugacity $f=e^{\frac{\mu}{kT}}=\frac{1}{A}$ and the chemical potential $\mu=-B\ln A$ (see Sect.~\ref{theory}) of both the original and randomized versions of the text ``Senilità''. If we look at the data, we realize at once the effect of word randomization. Indeed, we notice that, in the ``Senilità'' case, there has been an increase in the total energy $E$ of more than three times (from $60,296,678$  to $225,613,236$, see above). More important, the chemical potential $\mu$ has changed from $-4.4073$ to $-3,829.26$. This is mainly due to the fact that the constant $B$ in Eq. (\ref{be}) has drastically changed from $8,713.51$ to $55,935.53$, i.e. more than six times. In physics, the constant $B$ expresses the temperature in energy units or, better, the heat energy. In other words, randomizing the words of the original text has dramatically increased the `heat' of the text. These results suggest that, in the ``Senilità'' case, the effect produced by word randomization can be considered as a possible linguistic-conceptual counterpart of the effect produced by an increase of the temperature in a physical system.  

However, together with the notion of temperature, also the physical notion of `excitation' should be introduced. Indeed, let us consider the most frequent word in the text ``Senilità'', namely, \emph{Di}. Its number of appearance is equal to $2,120$ in the original version of the text, while it drastically decreases to a value of $21$ in the randomized version.\footnote{We are here assuming that the rank of the different words, in the transition between the original text and the randomized text, is preserved, similar to what would happen in a physical system, where the transfer of heat only leads to a change in the occupation numbers of the different energy levels, without changing the latter.} 
This decrease can be interpreted as the linguistic-conceptual counterpart of a phenomenon of local excitation, i.e. transfer of kinetic energy provoked by heating, like the random bombardment provoked by a heath bath at room temperature (see also footnote \ref{weather}). This local random effect entails that some of the words of the original text get excited, in the sense that they jump to a higher energy level, while others remain in the same energy level. 
In particular, out of $2,120$ placeholders of the word \emph{Di} that were originally in the lowest energy level, only $21$ remain at this level, whereas the others get excited and jump to a higher energy level. And the same systematically happens with the other words of the text. In other words, we can describe the word randomization process also as a `word excitation' process. 

As anticipated at the beginning of the section, the effect of word randomization on the statistical distribution of the words in a text was the same in `all' literary texts of Italian language, namely, despite randomization, Bose--Einstein statistics is still a good model for empirical data, but the differences with Maxwell--Boltzmann statistics are much less remarkable now. Our hypothesis is that two aspects play a role in the explanation of this empirical pattern. On the one side, some meaning is still preserved and, moreover, identical words are still indistinguishable in the randomized text. On the other side, the original coherence in terms of meaning of the whole text is partly destroyed by the random effects produced by what we have described as a linguistic-conceptual counterpart of an increase of temperature, which makes words, hence micros-states of the cognitons, behave more independently.

The results above confirm and strengthen the results obtained on word randomization on texts of English language, including the texts sketched in Sect.~\ref{theory} \citep{independence2024}. As already seen in Sect.~\ref{experiment}, this seems to indicate that general meaning-related patterns can be detected in human language which are `species-specific', as they are independent of the specific language that is used to investigate them. As mentioned at the beginning of the section, we plan to complete the study of a thermodynamic theory of human language in a forthcoming article \citep{temperature2025}.

\section{Quantum statistics and meaning\label{explanation}}
We have recently provided an explanatory hypothesis on the role that meaning might play in human language \citep{aertsbeltran2020,aertsbeltran2022b,independence2024}. This hypothesis seems corroborated by the results obtained in this article. Thus, let us dedicated this section to illustrate how meaning is connected with the appearance of quantum statistics in linguistic-conceptual domains.

The way in which meaning is attributed in the combination, or composition, of two concepts, i.e. two conceptual entities in our quantum theoretical framework (see Sect.~\ref{theory}), closely resembles the formation of entanglement in the composition, or combination, of two physical entities. In both cases, indeed, a phenomenon of `contextual updating' occurs, which we have recently formulated \citep{independence2024,philtransa2023,ijtp2023}. Thus, let us illustrate what we mean by contextual updating and firstly consider the combination of two words, and the two corresponding concepts, in a piece of text produced by human language.  

Each of the two words carries meaning but, whenever they are combined in a piece of text, their combination carries meaning in itself, which is not generally related to the original meanings of the two words in the way that is prescribed by a classical compositional semantics. In most cases, indeed,    `new emergent meaning' arises as a result of the combination. As a simple example, consider the two words \emph{White} and \emph{Lie}. The first often refers to the color or metaphorically to purity or innocence, the second to a false statement. However, the combination \emph{White lie} does not mean a lie that is literally white, or a lie about purity. Instead, it conveys a new meaning, that of a harmless or trivial lie told to avoid hurting someone's feelings. Hence, the combined meaning arises from a contextual interpretation of the two words. And the same happens when a word is added to an entire piece of text. This means that, each time a new word is added to a piece of text, a phenomenon of updating of meaning occurs which is influenced by the meaning of the whole context, and this contextual updating continues to occur until all words are added and the entire text is finalized.

If one represents mathematically the phenomenon of contextual updating, one discovers that meaning is attributed through the formation of quantum entanglement. Indeed, in the quantum theoretical framework presented in Sect.~\ref{theory}, the conceptual entities corresponding to the two words are associated with two Hilbert spaces and their combination, or composition, is associated with the tensor product Hilbert space. In the combination process, new states form as a consequence of the superposition principle, which always contains a majority of entangled states. These entangled states exactly accomplish the phenomenon of contextual updating in the quantum formalism.  

Hence, the phenomenon of contextual updating can explain the way in which meaning attaches to and intertwines with the words in the composition of a piece of text. Moreover, in the process of writing, the writer has typically the intention of making the piece of text as clear as possible. 
In our quantum theoretical framework, this corresponds to the preparation of a pure entangled state, whose von Neumann entropy is equal to zero. Thus, in the writing up process, the words, thus the conceptual entities corresponding to them, always entangle together, namely, `collaborate' with the aim of reducing the von Neumann entropy, hence, the overall uncertainty of the text, and producing globally a pure state \citep{philtransa2023,ijtp2023}. More specifically, in this collaborative dynamics to produce entanglement, identical words, i.e. cognitons in the same micro-states, tend to clump, or cluster, together, thus breaking the statistical independence. It is this dynamics of meaning formation that would be responsible of the appearance of 
the quantum Bose--Einstein statistics, in human language.  

\begin{scriptsize}
\begin{table}
\begin{center}
\begin{tabular}{ |c|c|c|c|c|c|c| } 
\hline 
Text & \multicolumn{3}{|c}{Original version} & \multicolumn{3}{|c|}{Randomized version} \\
  & Tot. energy & Fugacity & Chem. pot. & Tot. energy & Fugacity & Chem. pot. \\   
\hline 
${\cal T}$ & $E$ & $f
=\frac{1}{A}$ & $\mu=-B\ln A$ & $E$ & $f
=\frac{1}{A}$ & $\mu=-B\ln A$ \\
\hline
RS & 7,706,035 & 0.9986 & --2.9141 & 19,291,372 & 0.9202 & --625.22 \\
Cu & 83,981,879 & 0.9996 & --4.1283 & 323,987,875 & 0.9362 & --4,608.00 \\
Pi & 28,022,286 & 0.9989 & --6.7476 & 91,322,097 & 0.9295 & --2,382.55 \\
Se & 60,296,678 & 0.9995 & --4.4073 & 225,613,236 & 0.9338 & --3,829.26 \\
JO & 25,544,173 & 0.9992 & --2.8646 & 68,821,224 & 0.9208 & --1,161.80 \\
\hline 
\end{tabular}
\caption{Table 3. We report the total energy of words $E$, the fugacity $f$ and the chemical potential $\mu$ of all five literary texts of Italian language analysed in Sect.~\ref{experiment}, in both the original and randomized version. As we can see, in all texts, while the fugacity is substantially unchanged, there has been an important increase in the total energy and the magnitude of the chemical potential, in the passage from the original to the randomized version. This clearly indicates that the effect of word randomized has determined an (effect corresponding to an) increase in the `temperature' of the text. \label{table3}}
\end{center}
\end{table}
\end{scriptsize}

Regarding the introduction of an `ad hoc' randomness, altering in a genuinely casual way the numbers of word appearance in the text, hence also the way that words combine, our explanation is the following. In this case, we have seen in Sect.~\ref{randomization} that Bose--Einstein statistics still provides a good model for the statistical distribution of the words, which means that the phenomenon of clumping, or clustering, in the same micro-states is still present, as the words carry meaning in themselves and all words in the original text are still used, albeit with relative frequencies modified by the randomisation process. However, the Bose--Einstein behaviour is magnified when the relative frequencies of the different words are exactly those required by the meaning the writer intended to convey when s/he wrote the story that the text narrates, thus realizing a pure state of the overall textual entity. 

We can also observe that a Bose--Einstein condensate would be a limit case where all words are in the same micro-state. Consider, e.g., the simple piece of text \emph{No, no, no, \ldots, no},  that is, a repetition of the word \emph{No} shouted by a person witnessing a toddler about to touch a hot stove. This is an example of a Bose-Einstein condensate, with all cognitons in one energy micro-state, forming a piece of text that is contextually closed in terms of meaning in which the repetition of the same word is used to create intensity, in this case in the attempt to stop the child before harm occurs. In fact, the overall energy of a word in a text is given by the number of times it appears in it, i.e. how populated the corresponding energy level is. If, then, all the cognitons forming a text are concentrated on the same energy level, the energy will be maximised on the meaning it conveys, giving that meaning greater communicative impact.

The above suggests that quantum coherence can be regarded as a signature of the presence of meaning in a text.\footnote{However, describing the quantum coherence of a collection of cognitons by only considering their statistical behaviour is not sufficient to completely characterise the meaning contained in the text they form. Indeed, by altering the positioning of the words that make up such text, it is possible to go from a text that is perfectly intelligible and meaningful to one that will appear difficult for the reader to understand, while leaving untouched its statistical description. In this sense, the observation that the words of a text distribute according to Bose--Einstein statistics only reveals the presence of meaning at the potential level, in the sense that we know that these words can be used to create a meaningful story (which not all word collections allow doing), whereas of course they can also be used to create an almost unintelligible text.} 
Moreover, the introduction of randomness in these linguistic-conceptual domains could play for meaning the decoherence role that an increase of the temperature provokes in physical domains, namely, a local excitation effect which makes some words jump from lower to higher energy levels, destroying in this way the original meaning coherence, disentangling words and making them behave more independently. This would explain why Maxwell--Boltzmann statistics starts to appear in regimes where randomization is introduced. We do not deal in detail here with this phenomenon, which we however consider as crucial to understand the role of meaning and coherence, for the sake of brevity (see again \citet{temperature2025}).

\section{Quantum statistics and physics\label{physics}}
In the early days of quantum mechanics, the founding fathers of the theory did not attribute the appearance of quantum statistics, in particular, the Bose--Einstein statistics, to the complete physical indistinguishability of identical quantum entities. They rather connected a statistical behaviour that deviated from the Maxwell--Boltzmann statistics, e.g., that detected in a gas of photons, to the lack of statistical independence of the micro-states of physical entities (see footnote \ref{statind}). In particular, Albert Einstein was concerned that identical photons could behave in a statistically dependent way and wondered about the existence of a `mysterious force, whose nature is unknown, that makes them attract each other in an ideal gas' (for a detailed historical reconstruction of the scientific debate leading to the discovery of Bose--Einstein statistics, see, e.g., \citet{howard1990,darrigol1991,monaldi2009,perezsauer2010,gorroochurn2018}). 

With the advent of quantum mechanics in 1925--26, the problem of the statistical dependence of identical physical entities disappeared, because a new postulate, namely, the symmetry exchange postulate, was introduced by law, which forces the mathematics of Hilbert spaces to entangled (symmetric or anti-symmetric) state vectors and automatically incorporates the above mentioned lack of statistical independence. We have however anticipated in Sect.~\ref{intro} that the symmetry exchange postulate raises both epistemological and physical issues. From an epistemological point of view, the request of a defined symmetry in the state vector of two identical quantum entities would forbid the possibility of preparing a state of the composite entity in which the component entities are well localized in space and only entangled in spin \citep{krause2010,diekslubberdink2020}. From a physical point of view, the physical indistinguishability, theoretically incorporated into the symmetry exchange postulate, would regard identical entities with different energies, e.g., a blue photon and a red photon, as completely indistinguishable. This is at odd with the fact that quantum experimentalists treat as distinguishable two photons with different energies when they are produced by different sources, as is the case in their linear optics experiments for the creation of entanglement.

In regard to the latter point, one of the techniques to create entanglement for quantum information and computation tasks involves using photons of different color, i.e. different frequency and energy (see, e.g., \citet{zhao2014,fedrizzi2009,patel2010}). In one of these experiments \citep{zhao2014}, physicists managed to entangle photons coming from different sources and having different color. More precisely, two photons are sent into a beam splitter and entanglement is produced through a phenomenon of two-photon interference. However, contrary to the received view of identical quantum entities, according to which photons would be completely indistinguishable, photons with different color do not need to be indistinguishable when they hit the beam splitter, but only when they are actually measured. Before being measured, the photons may have different color, and one can still create entanglement if the measuring apparatus is insensitive to color or the color information is erased before the measurement.  

That photons with different energy are regarded as not experimentally indistinguishable straightly converges towards our quantum theoretical framework. Indeed, we have seen in Sects. \ref{theory} and \ref{experiment} that different words, i.e. cognitons in different energy states, are distinguishable. 
But this is not the only insight into physics coming from our quantum theoretical framework. Indeed, the explanatory hypothesis put forward in Sect.~\ref{explanation} provides new insights towards the identification of a physically more grounded justification for the appearance of the Bose--Einstein statistic in physics than that of the symmetry exchange postulate \citep{independence2024}. Indeed, we believe that, similarly to linguistic-conceptual domains, also in physics domains the lack of statistical independence would be produced by a phenomenon of entanglement through contextual updating. We clarify the sense of this statement in the following.

When an electron is added to an atom, the interaction with the other electrons already present in the atom and with its nucleus, makes the electron 
become entangled with them. Analogously, when a photon is added to a gas, the photon entangles with the other photons in the gas, through mutual interactions, which exactly occurs through the above mentioned phenomenon of contextual updating. Moreover, all the photons composing the gas collaborate to reduce the overall uncertainty, so that the gas is in a pure entangled state, i.e. its von Neumann entropy is zero, preserving in this way the overall quantum coherence. This would explain the tendency of identical photons and, more generally, of bosons, to clump, or cluster, together in the same micro-states. It would occur, as we said, through a phenomenon of contextual updating. This collective behaviour reaches its limit in the Bose--Einstein condensate at very low temperatures, where the available energy is so low that all bosons are in the same micro-state, that is, the minimal energy state. In this regard, an increase of the temperature would start destroying the overall coherence of the gas, until a point where the individual photons behave as if they were statistically independent. But, strictly speaking, there is no physical situation in which the bosons behave in a truly random and independent manner. This contrasts the general tenet of the classical statistical interpretation of thermodynamics, according to which physical entities behave in a random and independent way at the micro-level, which entails maximization of entropy \citep{huang1987}. On the contrary, a collaborative 
 dynamics is always present, whereby identical bosons work together in an attempt to reduce the overall entropy of the gas, within the limits of thermal bombardment. 

We think that the dynamics of entanglement formation through contextual updating, on the one hand provides a more convincing physical justification for the appearance of quantum mechanical statistics in physics and, on the other hand accords with Einstein's idea of an attractive force linking identical photons in a gas. This force would be the physics counterpart of a `force of meaning'.

\section{Conclusion \label{conclusion}}
We conclude this article with two remarks, as follows.Firstly, the results obtained in our investigation of quantum cognition suggest the hypothesis that the role played by the human mind with respect to language is the same as the role played by the context (measuring apparatus, heat bath, etc.) with respect to quantum physical entities. This is at the basis of a novel interpretation of quantum mechanics, the `conceptuality interpretation', initially put forward by one of the authors, which provides new interesting solutions to many of the long-standing quantum mysteries \citep{aertsetal2020,ass2025a,ass2025b}. Secondly, it is our opinion that our quantum theoretical approach to human language also provides a novel theoretical direction for the investigation of a thermodynamics of quantum physical entities, whereas existing approaches, mainly based on the quantum theory of open systems are not generally accepted by the scientific community (see, e.g.,  \citet{gemmeretal2009,mahler2015,gogolineisert2016}).

\section*{Acknowledgements}
This work was supported by the project ``New Methodologies for Information Access and Retrieval with Applications to the Digital Humanities'', scientist in charge S. Sozzo, financed within the fund ``DIUM -- Department of Excellence 2023--27'' and by the funds that remained at the Vrije Universiteit Brussel at the completion of the ``QUARTZ (Quantum Information Access and Retrieval Theory)'' project, part of the ``Marie Sklodowska-Curie Innovative Training Network 721321'' of the ``European Unions Horizon 2020'' research and innovation program, with Diederik Aerts as principle investigator for the Brussels part of the network.

\end{document}